\documentclass{JHEP3}
\usepackage{epsfig,multicol,bbm,latexsym}

\title{Constraints on the dark matter annihilation scenario of
Fermi 130 GeV $\gamma$-ray line emission by continuous gamma-rays,
Milky Way halo, galaxy clusters and dwarf galaxies observations}

\author{Xiaoyuan Huang$^a$, Qiang Yuan$^{b,c}$, Peng-Fei Yin$^b$,
Xiao-Jun Bi$^b$, Xuelei Chen$^a$\\
$^a$National Astronomical Observatories, Chinese
Academy of Sciences, Beijing 100012, P. R. China \\
$^b$Key Laboratory of Particle Astrophysics, Institute of High Energy 
Physics, Chinese Academy of Sciences, Beijing 100049, P. R. China\\
$^c$Key Laboratory of Dark Matter and Space Astronomy, Purple Mountain 
Observatory, Chinese Academy of Sciences, Nanjing 210008, P. R. China\\
huangxiaoyuan@gmail.com,yuanq@ihep.ac.cn,yinpf@ihep.ac.cn,bixj@ihep.ac.cu,
xuelei@cosmology.bao.ac.cn}

\abstract{
It was recently reported that there may exist monochromatic
$\gamma$-ray emission at $\sim 130$ GeV from the Galactic center
in the Fermi Large Area Telescope data, which might be related
with dark matter (DM) annihilation. In this work we carry out a
comprehensive check of consistency of the results with the DM
annihilation scenario, using the $3.7$ yrs Fermi observation of the 
inner Galaxy, Galactic halo, clusters of galaxies and dwarf galaxies.
The results found are as follows. 1) Very strong constraints on the 
DM annihilation into continuous $\gamma$-rays from the Galactic center 
are set, which are as stringent as the ``natural'' scale assuming 
thermal freeze-out of DM. Such limit sets strong constraint on the 
DM models to explain the line emission. 2) No line emission from the 
Galactic halo is found in the Fermi data, and the constraints on line
emission is marginally consistent with the DM annihilation interpretation 
of the $\sim 130$ GeV line emission from the inner Galaxy. 3) No line 
emission from galaxy clusters and dwarf galaxies is detected, although 
possible concentration of photons from clusters in $120-140$ GeV is 
revealed. The constraints from clusters and dwarf galaxies are weak and 
consistent with the DM annihilation scenario to explain the $\sim 130$ 
GeV line emission.}


\preprint{1208.0267}

\begin{document}

\section{Introduction}

Through analyzing the public Fermi Large Area Telescope
(Fermi-LAT) data, several groups reported the hints of
monochromatic $\gamma$-ray emission with energy $E\approx130$ GeV
\cite{2012JCAP...07..054B,2012JCAP...08..007W,2012JCAP...09..032T,
2012arXiv1205.4700B,2012arXiv1206.1616S}. The morphology of
the potential line emission is still in debate. In some works the
results showed that there were various regions lying basically in
the Galactic plane having excess in $120-140$ GeV
\cite{2012JCAP...09..032T,2012arXiv1205.4700B}, while in another work 
the excess concentrated in the inner $5^{\circ}$ of the Milky Way center
\cite{2012arXiv1206.1616S}. The results stimulated active discussions of the
possible dark matter (DM) origin \cite{2012arXiv1205.1520D,
2012PhRvD..86a5016C,2012PhRvD..86d3515C,2012arXiv1205.4151K,
2012arXiv1205.4675L,2012JCAP...09..003R,2012arXiv1205.5789S,
2012PhRvD..86h3521C,2012arXiv1206.2910W,2012arXiv1206.4758F,
2012arXiv1207.1341H,2012arXiv1207.1537O,2012PhLB..715..285Y,
2012JCAP...10..033F,2012arXiv1207.4981P,2012JCAP...08..003D,
2012arXiv1206.2863K}. Alternatively the astrophysical explanations of 
the line-like $\gamma$-ray emission were also proposed 
\cite{2012JCAP...07..011P,2012arXiv1207.0458A}.

Due to the importance of the high energy line emission, many works
were trying to test the line emission (and its DM interpretation)
with Fermi-LAT observations of continuous $\gamma$-rays
\cite{2012PhRvD..86d3524B,2012arXiv1206.7056B,2012arXiv1207.0800C,
2012PhRvD..86h3525C}, the line emission from dwarf galaxies and clusters 
of galaxies \cite{2012PhRvD..86b1302G,2012arXiv1207.4466H}, 
the unassociated Fermi sources \cite{2012arXiv1207.7060S}, and the 
future detection with high energy resolution telescope 
\cite{2012PhLB..715...35L,2012arXiv1207.6773B}. The basic results 
show that the allowed cross section of the DM annihilation
final states giving rise to the continuous emission can only be
about $O(10^2)$ times larger than that of the line emission
\cite{2012PhRvD..86d3524B,2012arXiv1206.7056B,2012arXiv1207.0800C,
2012PhRvD..86h3525C}, which needs to be considered when constructing 
the DM models. The recent search for line emission in the Milky Way 
halo by Fermi collaboration with two-year PASS 6 data showed no indication 
of signals \cite{2012PhRvD..86b2002A}, and the upper limits seemed 
to be marginally in conflict with the results found in 
\cite{2012JCAP...08..007W,2012JCAP...09..032T}.
Furthermore, no signal from dwarf galaxies was found and the
constraints on DM annihilation into monochromatic $\gamma$-rays
were consistent with the results found in the inner Galaxy 
\cite{2012PhRvD..86b1302G}. For the galaxy clusters, however, it was 
claimed a $\sim3\sigma$ signal of possible $130$ GeV line emission 
by Hektor et al. \cite{2012arXiv1207.4466H}. 

In this work we try to make a comprehensive test of the DM
scenario of the line emission, through analyzing the Fermi-LAT
$\gamma$-ray data in the Galactic center, the Milky Way halo,
dwarf galaxies and galaxy clusters. In our analysis the
information of the spatial distribution of DM is taken into
account. We further include the effects of substructures on the DM
annihilation in the analysis, both the enhancement of annihilation
luminosity and the change of surface brightness distribution
\cite{2012MNRAS.419.1721G,2012MNRAS.425.2169G}.

This paper is organized as follows. In Sec. II we will discuss the
constraints on the continuous $\gamma$-ray emission from the Galactic
center region. In Sec. III, IV and V we give the results of the line
search from the Milky Way halo, galaxy clusters and dwarf galaxies
respectively. Finally Sec. VI is our conclusion and discussion.

\section{Continuous gamma-ray emission from the inner Galaxy}

As DM is a neutral particle it can not couple to photons directly.
Therefore DM annihilate into to monochromatic photons through loop
processes with charged virtual particles in the loop. If DM annihilate 
into the charged particles at the tree level, such as quarks, charged 
leptons or gauge bosons, these annihilation products may induce 
significant continuous $\gamma$-ray flux. Using the Fermi data we can 
set a strong constraint on the DM annihilation cross section into these 
final states.

The current most stringent limits on the DM annihilation cross
section set by the Fermi collaboration were derived from the analysis 
of ten dwarf satellite galaxies with 24 months of Fermi data 
\cite{2011PhRvL.107x1302A}. The Galactic center (GC) has been thought 
to be the best target for probing the DM annihilation signals. It may 
have larger DM annihilation rate than dwarf galaxies, and is a good
target for searching for the continuous $\gamma$-rays induced by
DM. In this section, we will use the Fermi data from the inner
Galaxy to place limits on the DM annihilation cross section. This
analysis is important for building the DM model to explain the
monochromatic $\gamma$-ray emission as it has to satisfy the
continuous $\gamma$-ray limit simultaneously.

In this analysis we use the $3.7$ years Fermi-LAT
data\footnote{http://fermi.gsfc.nasa.gov/ssc/data} recorded from 4
August 2008 to 18 April 2012, with the Pass 7 photon selection.
The ``SOURCE'' (evclass=2) event class is selected, and the
recommended filter cut ``(DATA\_QUAL==1) \&\& (LAT\_CONFIG==1)
\&\& ABS(ROCK\_ANGLE)$<52$'' is applied. The energy range of
events is restricted from $500$ MeV to $300$ GeV, and the
region-of-interest (ROI) is adopted to be a
$10^{\circ}\times10^{\circ}$ box centered around the GC. We carry
out the binned likelihood analysis with the LAT Scientific Tools
v9r23p1. The instrument response function used is ``{\tt
P7SOURCE\_V6}''. For the diffuse background, we use the Galactic
diffuse model {\tt gal\_2yearp7v6\_v0.fits} and the isotropic
diffuse spectrum {\tt iso\_p7v6source.txt} provided by the Fermi
Science Support
Center\footnote{http://fermi.gsfc.nasa.gov/ssc/data/access/lat/Background-
Models.html}. The Galactic diffuse model is based on the interstellar 
gas distribution from spectral line surveys and the dust distribution 
from infrared observations. An inverse Compton scattering component 
based on GALPROP \cite{1998ApJ...509..212S} was also included in the 
modelling. Together with the isotropic diffuse background, the model 
template was derived through fitting the LAT diffuse $\gamma$-ray data.

The data are binned with $30$ energies bins logarithmically spaced
and $100\times100$ spatial bins with size $0.1^{\circ}$ each. For
the likelihood analysis we include the point sources within
$5^{\circ}$ around the Galactic center based on the 2-year LAT
source catalog (using the user-contributed software
make2FGLxml.py) \cite{2012ApJS..199...31N}. For other sources located in 
the ROI we fix their parameters to be the values of the LAT source
catalog. Before taking the DM contribution into account, we first
make a global fit to derive the spectral parameters of the point
sources. All the spectral parameters of the sources within
$5^{\circ}$ of the Galactic center and the normalizations of the
diffuse backgrounds are left free during the fit. Then we add the
DM component as a diffuse source, and re-do the fit to get the
upper limits of the DM contribution. The free parameters of the
latter fit include the normalizations of the point sources within
$5^{\circ}$ of the Galactic center, the normalizations of the
diffuse backgrounds and the DM component.

For the DM component we discuss $W^+W^-$, $b\bar{b}$, $\mu^+\mu^-$
and $\tau^+\tau^-$ final states. The photon production spectra are
calculated with the PYTHIA simulation package (version 6.4, 
\cite{2006JHEP...05..026S}). The spatial density distribution of DM 
is assumed to be either an Navarro-Frenk-White (NFW, 
\cite{1997ApJ...490..493N}) profile
\begin{equation}
\rho(r)=\frac{\rho_s}{(r/r_s)(1+r/r_s)^2},
\end{equation}
with $r_s=20$ kpc and $\rho_s=0.35$ GeV cm$^{-3}$, or an Einasto
profile \cite{1965TrAlm...5...87E}
\begin{equation}
\rho(r)=\rho_{\rm s}\exp\left(-\frac{2}{\alpha}\left[\left(\frac{r}
{r_{\rm s}}\right)^{\alpha}-1\right]\right),
\end{equation}
where $\alpha=0.17$, $r_s=20$ kpc and $\rho_s=0.08$ GeV cm$^{-3}$.
The scale density of both the profiles is to give a local DM density
$\simeq0.4$ GeV cm$^{-3}$ \cite{2010JCAP...08..004C}.

In the analysis we also take the DM substructures into account. We
employ the results from Aquarius simulations \cite{2008MNRAS.391.1685S} 
and the fitted annihilation rate from the substructures in Ref.
\cite{2012arXiv1203.5636Y}
\begin{eqnarray}
\langle\rho_{\rm sub}\rangle^2&=&9.3\times10^{-4}\times\frac{M_{\rm min}^
{-1.14}-M_{\rm max}^{-1.14}}{M_{\rm res}^{-1.14}-M_{\rm max}^{-1.14}}\nonumber\\
&\times&\frac{1}{1+(r/54\,{\rm kpc})^{2.76}}\,{\rm GeV^2\,cm^{-6}},
\end{eqnarray}
where $M_{\rm max}\approx10^{10}$ M$_{\odot}$ is the maximum mass
of subhaloes in the Milky Way halo, $M_{\rm
res}\approx3\times10^5$ M$_{\odot}$ is the resolution mass of the
simulation, and the minimum subhalo mass is assumed to be $M_{\rm
min}\approx10^{-6}$ M$_{\odot}$. For the inner Galaxy discussed in
this section, the substructures play little role in the
enhancement of the annihilation signal.

We project the density square into a 2-dimensional surface map to give
the spatial template of the DM annihilation. With the spatial template
and the energy spectrum, we use the above described likelihood analysis
method to derive the upper limits of the DM component. The only parameter
of the DM component is the normalization factor.

\FIGURE{
\includegraphics[width=0.6\columnwidth]{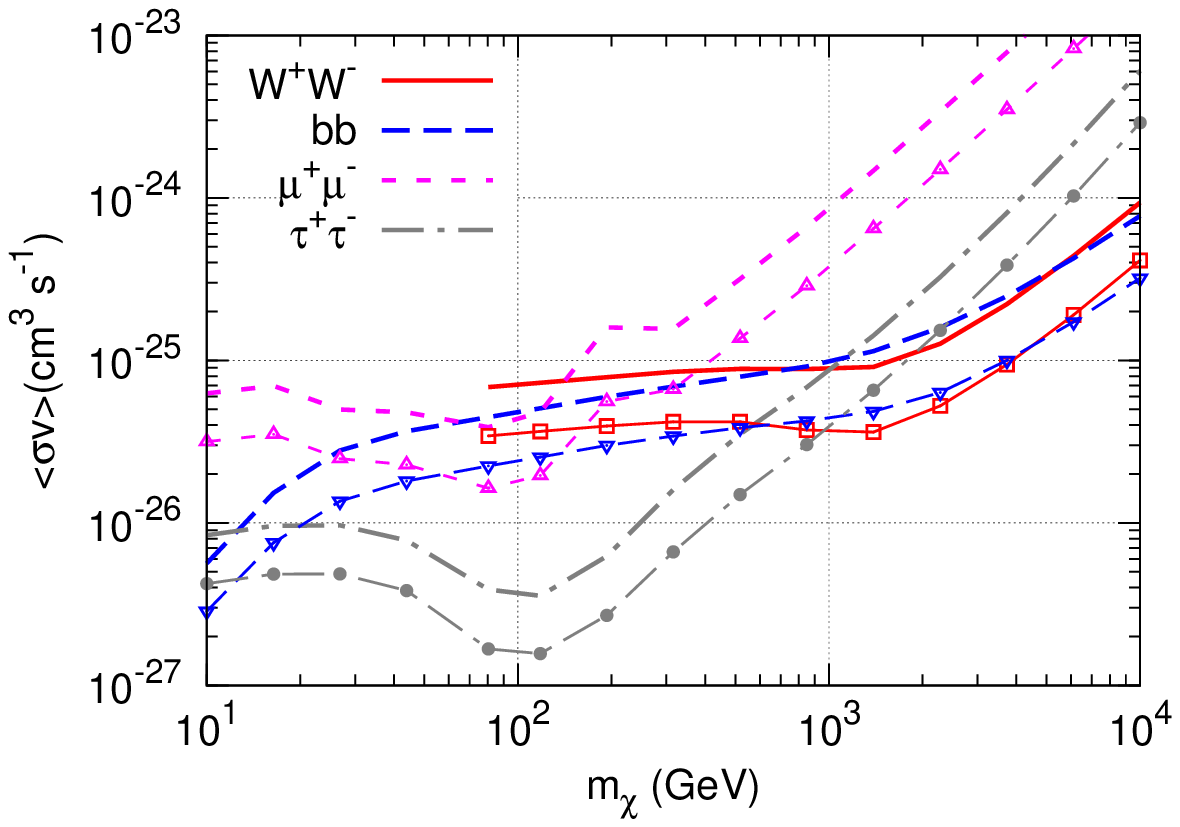}
\caption{$95\%$ confidence level constraints on the DM annihilation 
cross sections to $W^+W^-$ (solid), $b\bar{b}$ (dashed), $\mu^+\mu^-$ 
(short-dashed) and $\tau^+\tau^-$ (dash-dotted), for $10^{\circ}\times
10^{\circ}$ region around the Galactic center. Lines are for NFW profile, 
and lines with points are for Einasto profile.
\label{fig:sv_gc10}}
}

The results of the constraints are shown in Figure
\ref{fig:sv_gc10}. It can be seen that for DM particle with mass
less than hundreds of GeV most of the constraints can reach the
``natural'' scale, $3\times 10^{-26}$ cm$^3$ s$^{-1}$, assuming DM
is the thermal relic of the early universe. These constraints are
so far the most stringent constraints on the DM annihilation with
Fermi-LAT $\gamma$-ray data. Compared with \cite{2012PhRvD..86b2002A}, 
our constraints are more stringent because we use the likelihood fit
involving the astrophysical background, instead of assuming the DM
component not to exceed the observational data conservatively.
Also in \cite{2011PhRvD..84l3005H} the constraints were given directly 
by comparing the DM contribution with the observed excess in the
Galactic center region with the known sources and diffuse emission
subtracted.

For DM mass $\sim130$ GeV and NFW (Einasto) profile, the
constraints on the cross sections of $W^+W^-$, $b\bar{b}$,
$\mu^+\mu^-$ and $\tau^+\tau^-$ are $7.4\times10^{-26}$
($3.7\times10^{-26}$), $5.1\times10^{-26}$ ($2.6\times10^{-26}$),
$4.8\times10^{-26}$ ($2.0\times10^{-26}$) and $3.6\times10^{-27}$
($1.6\times10^{-27}$) cm$^3$ s$^{-1}$ respectively. Compared with
the cross section to monochromatic $\gamma$-rays $\langle\sigma v
\rangle_{\rm mono} \approx2.3\times10^{-27}$ ($1.3\times10^{-27}$) 
cm$^3$ s$^{-1}$ \cite{2012JCAP...08..007W}, the continuous cross section 
can only be larger by a factor of $1.2-30$. Our results are stronger 
than that derived in \cite{2012arXiv1206.7056B,2012PhRvD..86h3525C}, and 
are comparable with that in \cite{2012arXiv1207.0800C}.

Before the discussion of the physical implication of such constraints,
we need to be cautious about the possible systematic uncertainties of the
constraints. The discussion of systematic uncertainties from the instrument
response functions and calibration can be found in \cite{2012ApJS..203....4A}.
Here we discuss a little bit about the systematic effects from the choice
made in this analysis. The adopted energy band may cause systematic 
uncertainties of the constraints. Since we discuss the DM particle with
mass as low as $10$ GeV, the lower limit of photon energies should not be
too high in order to give effective constraints. On the other hand to
avoid the influence of large resolution angle, and to save the computation 
time the low energy limit is better to be higher. Through varying the low 
energy limit of selected photons from $500$ MeV to $2$ GeV, we find the
derived upper limits will change by tens of percents for $m_{\chi}\lesssim
300$ GeV. For higher masses of DM, the differences become larger and at
most a factor of $\sim3$. This is primarily due to the differences of
the modeling of diffuse backgrounds for various energy bands. We further 
check that, through fixing or relaxing the point sources in the ROI, 
increasing or decreasing the ROI boxes, the derived constraints will 
change by a factor less than $2$. Thus the overall systematic uncertainties 
of the current results should be a factor of several.

Such strong constraints on the DM annihilation cross section of
the tree level processes are very important when constructing the
DM models to explain the monochromatic emission. In the following
we discuss the implications of the constraints briefly.

As we have mentioned, DM annihilate into monochromatic photons
through higher order processes with charged particles running in
the loop. These charged particles will induce large continues
$\gamma$-ray flux. Here we define a criterium as the ratio of the
tree level process into continuous $\gamma$-rays and the loop
process into line $\gamma$-ray, $R \equiv \langle\sigma v
\rangle_{\rm cont}/\langle\sigma v\rangle_{\rm mono} = 
\langle\sigma v\rangle_{X\bar{X}}/\langle\sigma v\rangle_{\gamma\gamma}$ 
where $X$ denote a charged particle. To satisfy the Fermi constraints, $R$ 
should be smaller than $1.2-30$, depending on different annihilation modes.

The charged particle, $X$, in the loop can be a SM particle and/or
an exotic particle in new physics. Since the loop processes are
generally suppressed, the large DM annihilation cross section into
monochromatic photons $\sim 10^{-27}$ cm$^3$ s$^{-1}$ often
requires additional enhancement mechanism. If all the charged
particles in the loop are SM particles, an additional narrow
resonance mediator $A$ is useful to enhance DM annihilation cross
section \cite{2012PhRvD..86d3524B,2012arXiv1206.2863K}. Using the 
narrow width approximation, $R$ can be determined by the branching 
ratios of $A$ decay $BR(A\to X\bar{X})/BR(A\to \gamma \gamma)$. In 
general, $R$ is proportional to $1/\alpha^2$ due to the loop factor and 
two QED couplings. If $X$ are quarks or leptons, $R$ would be enhanced
by an additional factor of $m_{\chi}^2/m_{X}^2$. In this case, the
continuous $\gamma$-ray flux is much larger than monochromatic
flux, for instance by a factor of $> 10^6$, and are certainly
excluded by Fermi results. If $X$ is $W$ boson, $R$ is $\sim
O(10^4)$ and can be excluded by continuous $\gamma$-ray
observations too.

If the mediated charged particles in the loop include both SM
particles and exotic particles in new physics, the estimation of
$R$ depends on the details of model, and is more complicated. A
familiar example is supersymmetry in which the neutralinos can
annihilate into photons through box loop including chargino-$W$
boson and sfermion-fermion contributions \cite{1994APh.....2..261B,
1995PhRvD..51.3121J,1997PhLB..411...86B,1997NuPhB.504...27B}. If the 
main component of DM is Wino or Higgsino, the diagram including
chargino-$W$ boson contributions would be dominated. In such
cases, the large Wino/Higgino-chargino-W couplings are also
helpful to obtain required annihilation cross section. However,
$R$ is still as large as $O(10^2)-O(10^3)$ at least
\cite{2012PhRvD..86h3525C}, and can be excluded by Fermi continuous
$\gamma$-ray limits.

Therefore, these constraints suggest the charged particles in the
loop have to be heavier than the DM particle so that the tree
level process into the charged particles is forbidden. For
example, a chargino triangle loop can be used to explain required
DM annihilation cross section into monochromatic photons
\cite{2012JCAP...08..003D,2012arXiv1206.2863K}. In this case, the 
processes producing line and continuum $\gamma$-ray spectra are 
independent. Therefore, no excess continuous $\gamma$-ray flux would 
be induced by adjusting model parameters.

Finally, we give another comment on the importance of the continuous 
$\gamma$-ray constraints. Any successful DM model needs to explain the 
observed DM relic density $\Omega h^2 \sim 0.11$. If DM is thermally 
produced through velocity independent annihilation process in the early 
universe, the cross section is about the ``natural'' scale 
$3\times 10^{-26}$ cm$^3$ s$^{-1}$. If the future upper-limits derived 
from continuous $\gamma$-ray observations on DM annihilation cross 
section are lower than the ``natural'' scale, it means the DM should 
be totally or partially produced through other mechanisms. For example, 
the corrected DM relic density can be obtained by co-annihilation, or 
velocity suppressed annihilation \cite{2012PhRvD..86d3524B}, or 
non-thermal production mechanism \cite{2012arXiv1203.5636Y}. These 
results will be important for the DM model building and DM detections.

\section{Line emission from the Galactic halo}

\FIGURE{
\includegraphics[width=0.45\columnwidth]{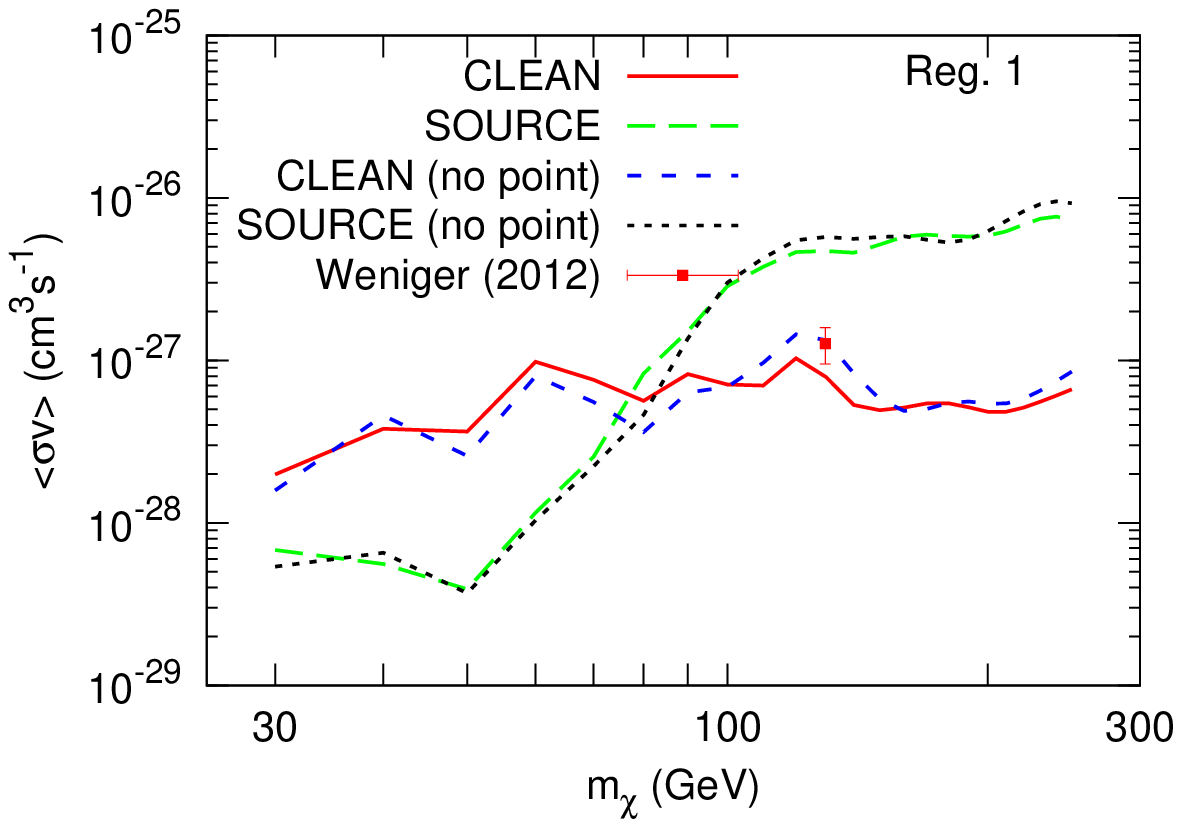}
\includegraphics[width=0.45\columnwidth]{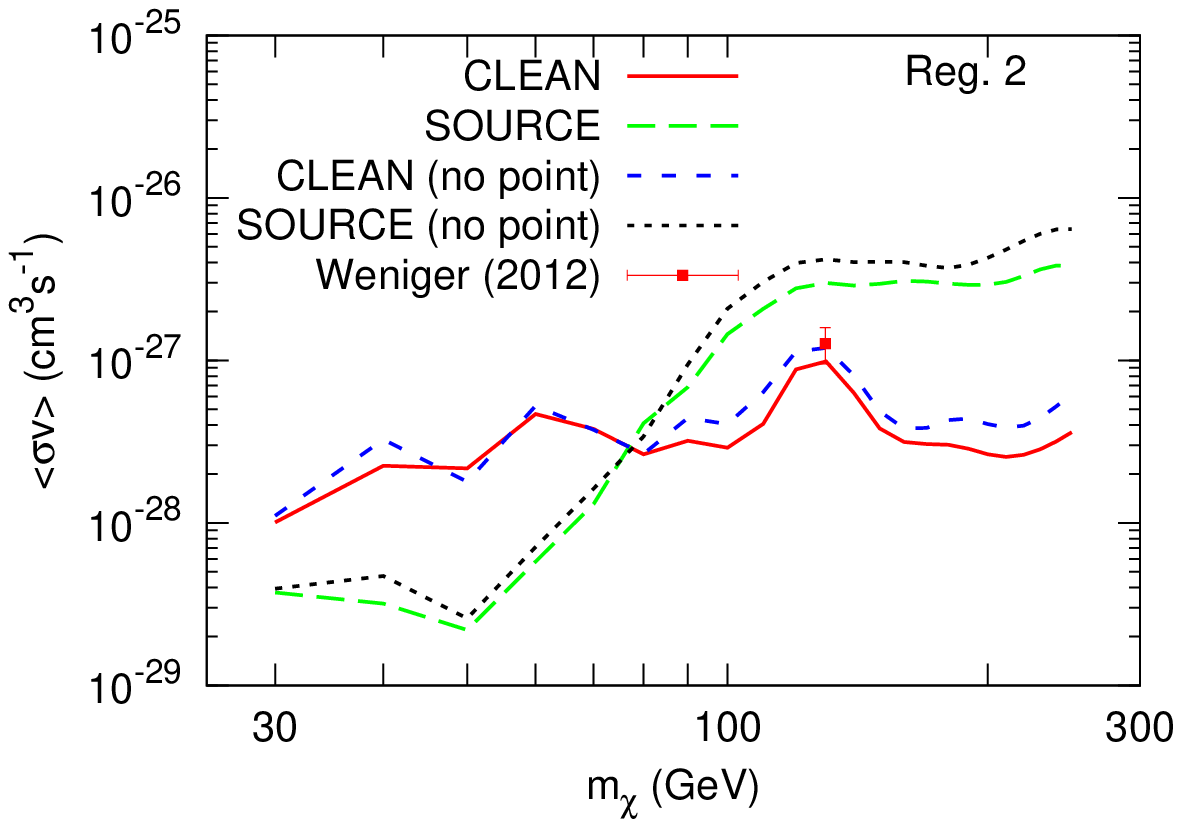}
\caption{Constrains on the DM annihilation cross section to
monochromatic $\gamma$-rays. Left panel is for Reg. 1 and right
panel is for Reg. 2. The DM density profile is assumed to be
Einasto profile, and no substructures are included.
\label{fig:sv_halo}}
}

\FIGURE{
\includegraphics[width=0.45\columnwidth]{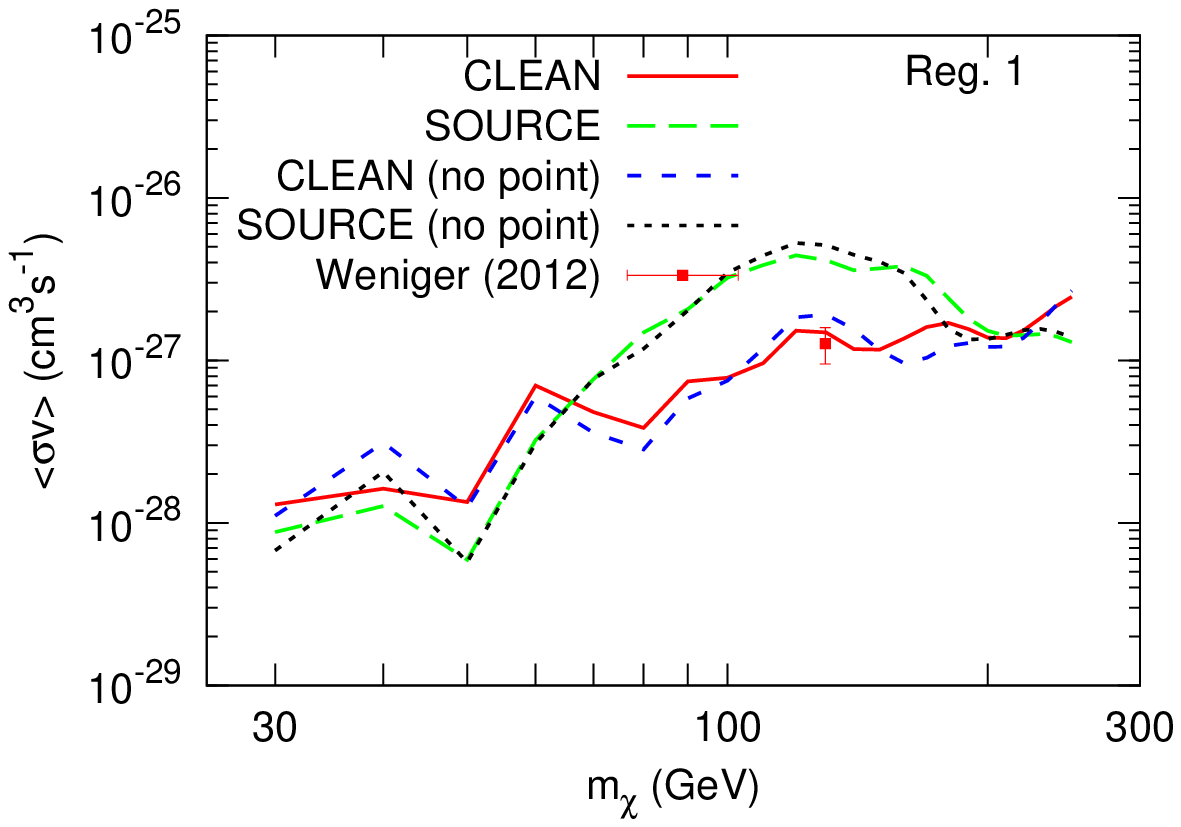}
\includegraphics[width=0.45\columnwidth]{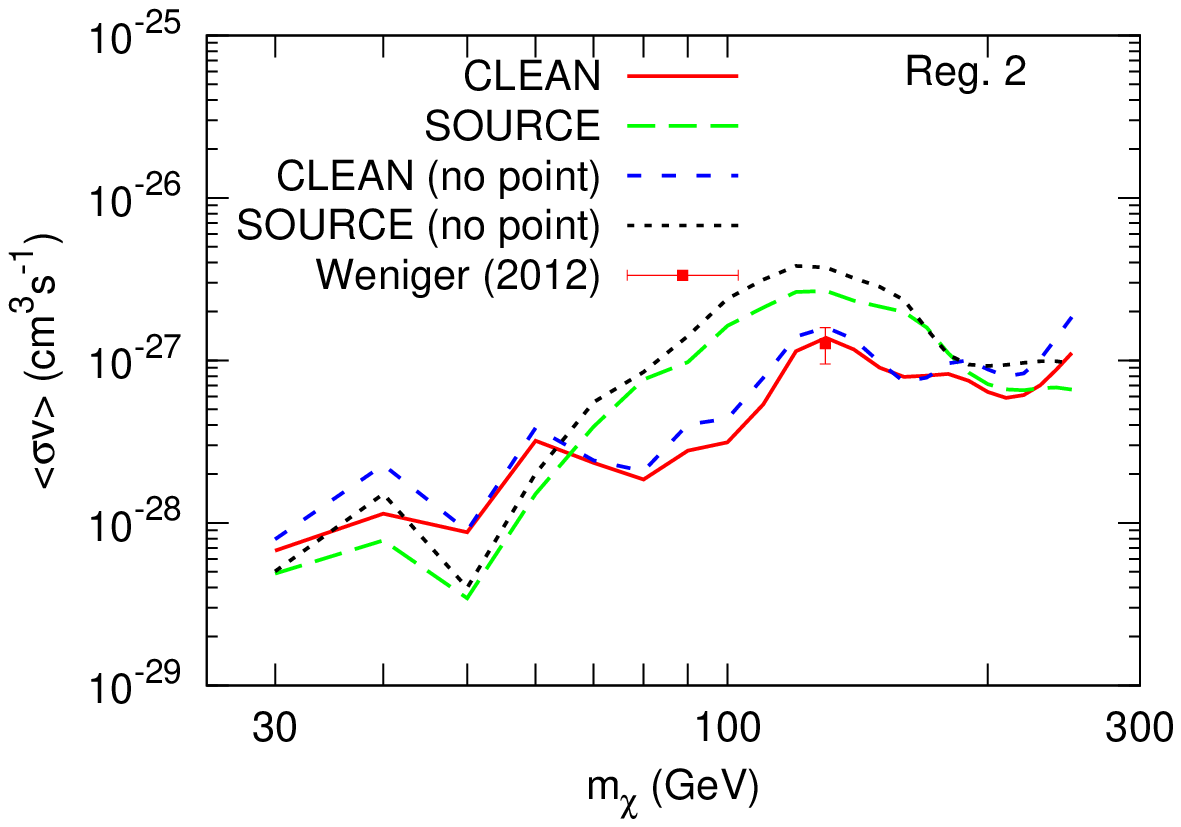}
\caption{Same as Figure \ref{fig:sv_halo} but for log-parabolic background
spectrum.
\label{fig:sv_halo2}}
}

The tentative $\gamma$-ray line emission was found mainly in the inner
Galaxy, which is expected in the DM annihilation scenario. As a
consistency check, the search for the line emission in the Galactic
halo should be of great importance, especially in case that there might
be significant contribution from DM substructures as shown by the
cold DM simulations. We use the same period and data version as in
Sec. II, but the whole sky region is adopted. Both the event classes
``SOURCE'' and ``CLEAN'' are analyzed. Since we focus on the high energy
line emission, the energy range in this analysis is adopted to be
$[20,\,300]$ GeV. We choose two sky regions in the analysis. The first
one (hereafter Reg. 1) is $|b|>10^{\circ}$, which can exclude most of
the Galactic plane and keep most of the halo region. The second one
(Reg. 2) is $|b|>10^{\circ}$ plus $|l|\leq10^{\circ},\,|b|\leq10^{\circ}$,
due to the fact that DM will concentrate in the central Galaxy and such
an adoption of the sky region may keep more potential DM signal. The latter
adoption is the same as that in Ref. \cite{2012PhRvD..86b2002A}. We also 
test the effect of point sources, through masking $1^{\circ}$ radius of 
each point source based on the second Fermi-LAT source catalog 
\cite{2012ApJS..199...31N}.

\TABLE{
\caption{$J_{\Delta\Omega}$ of the DM annihilation in the Milky Way halo.
In parenthesis are the results after masking point sources.}
\begin{tabular}{c|cccc}
\hline
\hline
       &  NFW & NFW+sub & EIN & EIN+sub \\
\hline
Reg. 1 & $20.2(17.7)$ & $43.9(38.7)$ & $22.0(19.3)$ & $45.8(40.3)$ \\
Reg. 2 & $31.4(22.5)$ & $55.4(43.6)$ & $42.2(27.8)$ & $66.2(48.9)$ \\
\hline
\hline
\end{tabular}
\label{table:jpsi}
}

To search for the $\gamma$-ray line emission, we perform a likelihood
fit to the photon counts using a continuous background together with a
monochromatic $\gamma$-ray line. The Possion likelihood function is 
defined as
\begin{equation}
\ln\mathcal{L}=\sum_i n_i\ln \phi_i-\phi_i-\ln n_i!,
\end{equation}
where $n_i$ is the observed photon counts in energy bin $i$, and $\phi_i$
is the expected counts (flux times exposure) in the energy bin. 
The continuous background spectrum in this energy range is 
approximated with a power-law or log-parabolic function. The energy 
resolution of Fermi-LAT detector (as given in \cite{2012PhRvD..86b2002A} 
with an average of the angular acceptance) has been taken into account 
when calculating $\phi_i$. The data in the energy range adopted here
are binned logarithmically into $400$ energy bins, which makes the
fit effectively ``unbinned'' \cite{2012JCAP...08..007W}. The $95\%$ 
confidence level upper limits on the line emission are derived with the 
profile likelihood method \cite{2005NIMPA.551..493R}, setting 
$-2\ln(\mathcal{L}/\mathcal{L}_{\rm best})
=2.71$, where $\mathcal{L}_{\rm best}$ is the best fit likelihood with
the line contribution. The photon upper limits can be translated into 
constraints on the DM annihilation cross section to monochromatic photons. 
To get the constraints on the cross section, we first compute the 
dimensionless astrophysical $J$-factor, defined as
\begin{equation}
J_{\Delta\Omega}=\frac{1}{\rho_{\odot}^2R_{\odot}}\int{\rm d}\Omega
\int_{\rm l.o.s.}\rho^2(l){\rm d}l,
\end{equation}
where $\rho_{\odot}=0.4$ GeV cm$^{-3}$ is the local density, $R_{\odot}
=8.5$ kpc is the distance from the Galactic center to the Earth, and
l.o.s. means integral along the line of sight. The $J$-factors for
NFW and Einasto profiles, with and without substructures are listed
in Table \ref{table:jpsi}. The numbers in parenthesis are the results
after masking $1^{\circ}$ radius around each point source of the 2-year
LAT source catalog \cite{2012ApJS..199...31N}. 

The $95\%$ confidence level constraints on the DM annihilation 
cross section are shown in Figures \ref{fig:sv_halo} and \ref{fig:sv_halo2}, 
for power-law and log-parabolic continuous background spectra respectively. 
The left panel of each figure is for Reg. 1 and the right panel is for 
Reg. 2. Here we adopt the Einasto density profile, and the possible 
effects from substructures are not considered. The cross section derived 
in Weniger (2012) \cite{2012JCAP...08..007W} is also shown for comparison. 
The differences between Figure \ref{fig:sv_halo} and Figure \ref{fig:sv_halo2} 
implies that the spectral shape for the continuous background is important 
for the search for line emission. For the ``SOURCE'' events, there is a 
residual cosmic ray component which makes the continuous background 
deviate from a single power-law. Therefore if assuming single power-law 
background, one may over-estimate (under-estimate) the background at 
the low (high) energy part and accordingly the constraints are stronger 
(weaker). For the ``CLEAN'' events we can see that the differences 
between these two continuous backgrounds are smaller, which means the 
background of the ``CLEAN'' events can be more or less approximated by 
a single power-law function. We also test the broken power-law spectrum
of the continuous background, which gives similar results with the
log-parabolic one.

The results show that the constraints from the Galactic halo are well
consistent with Weniger (2012) result from the Galactic center when we 
choose ``SOURCE'' data as done in \cite{2012JCAP...08..007W}. For the 
``CLEAN'' data the constraints are of the same level as the best-fit 
line-like ``signal'' \cite{2012JCAP...08..007W}. If the DM substructures 
are taken into account, the constraints from the Milky Way halo will be 
stronger by a factor of $\sim2$, which may then indicate a weak tension 
between the halo and the inner Galaxy. However, considering the 
uncertainties of the present constraints (from e.g., the adoption of 
continuous background), the errorbars of the fitting result of the 
``signal'', and the data sample used to derive the ``signal'' in 
\cite{2012JCAP...08..007W}, we can only conclude that the present 
constraints from the Milky Way halo observations of the line-like 
emission are marginally consistent with that from the inner Galaxy if 
explaining it with DM annihilation. The ``CLEAN'' results of Reg. 2 are 
also consistent with that derived by Fermi-LAT collaboration 
\cite{2012PhRvD..86b2002A}, in which two-year Pass 6 ``ULTRACLEAN'' 
data are used.

\section{Line emission from galaxy clusters}

Galaxy clusters (this section) and dwarf galaxies (next section) are
also potential good targets for probing DM signals. In this section we
analyze the Fermi-LAT data on a series of nearby clusters to search
for the potential line emission. The data version and time window are
the same as in the above discussions, but the energy range is set
from $200$ MeV to $300$ GeV. The ROI is adopted to be
$14^{\circ}\times14^{\circ}$ box centered in the target galaxy
cluster. Seven clusters, Fornax, AWM7, M49, NGC4636, Centaurus, Coma
and Virgo are choosen as our targets. The mass of Virgo is adopted
from \cite{2011PhRvD..84l3509P}, and for others the masses are 
adopted from the extended HIFLUGCS X-ray catalog 
\cite{2002ApJ...567..716R,2007A&A...466..805C}.

\FIGURE{
\includegraphics[width=0.45\columnwidth]{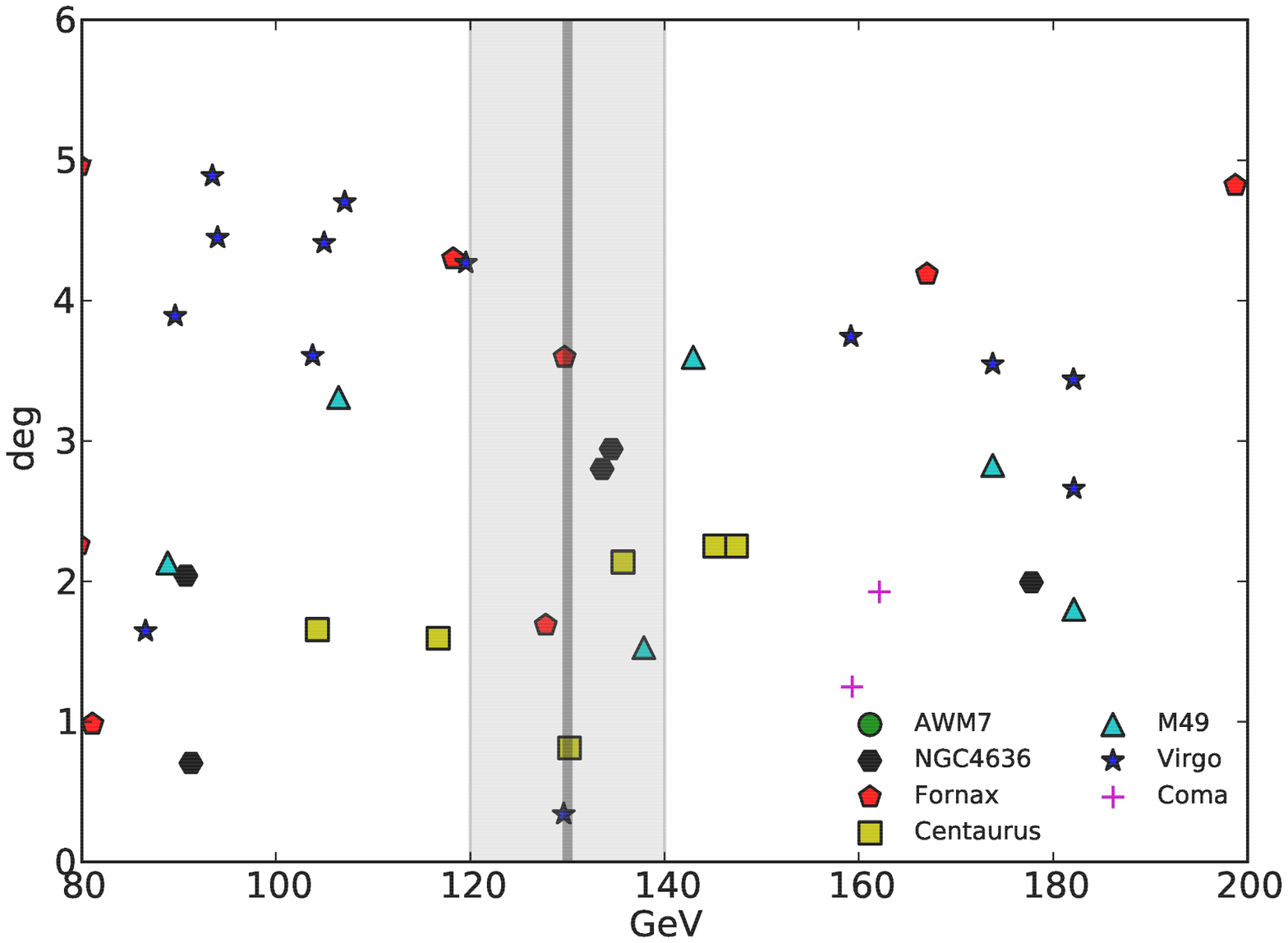}
\includegraphics[width=0.45\columnwidth]{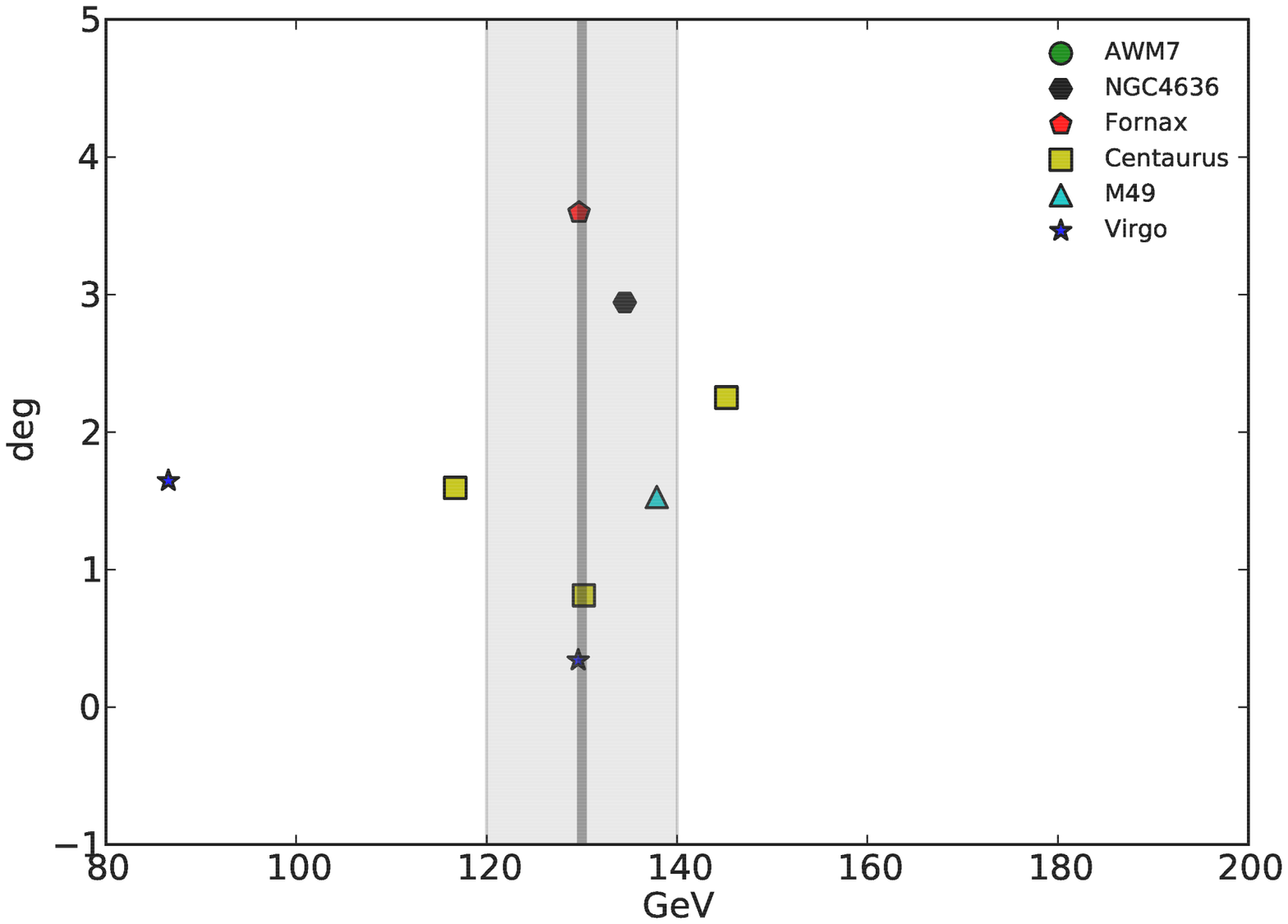}
\caption{Scattering plot of photon energies and deviation angles from
the cluster center, for the events within virial radius of each cluster.
Left panel is for ``SOURCE'' events, and right panel is for ``CLEAN''
events.
\label{fig:cnts}}
}

Figure \ref{fig:cnts} shows the individual photons with arrival
directions within the virial radius of these clusters. The left
panel is for ``SOURCE'' events, and the right panel is for
``CLEAN'' events. It is interesting to find that for the ``CLEAN''
events, photons with energies between $120$ and $140$ GeV are more
abundant than nearby energy ranges. We checked for the
``ULTRACLEAN'' events and find identical results with the
``CLEAN'' events. From Figure \ref{fig:cnts} it seems hard to claim
the existence of line emission. Recently it was found a $3\sigma$
evidence for the existence of $\sim130$ GeV line emission from
galaxy clusters in \cite{2012arXiv1207.4466H}.

We then derive the constraints on the DM annihilation cross
section to $\gamma$-ray line from the galaxy cluster observations.
The DM signals from galaxy clusters are modeled as extended
sources as in \cite{2012JCAP...01..042H}, with surface brightness
\begin{equation}
S_{\rm sm}(\theta)\propto\int_{\rm l.o.s.}\rho_{\rm sm}^2(l,\theta){\rm d}l,
\label{eqn:fluxADM}
\end{equation}
where $\theta$ is the angle between l.o.s. and the cluster center,
$\rho_{\rm sm}$ is the density profile of the smooth halo, which is
assumed to be NFW profile. The mass-concentration relation is adopted
to be \cite{2008MNRAS.390L..64D}
\begin{equation}
c(M_{200})=5.74\left(\frac{M_{200}}{2\times10^{12}h^{-1}M_{\odot}}
\right)^{-0.097}.
\end{equation}

The DM substructures may play a significant role on the enhancement
of annihilation signal from clusters \cite{2012MNRAS.419.1721G,
2012arXiv1201.1003H}. Following Ref. \cite{2012MNRAS.419.1721G} we adopt 
the boost factor of substructures as a function of cluster virial 
mass $M_{200}$
\begin{equation}
b(M_{200})=1.6\times10^{-3}(M_{200}/M_{\odot})^{0.39},
\end{equation}
and the surface brightness profile of subhalo emission
\begin{eqnarray}
S_{\rm sub}(\theta)&=&\frac{16b(M_{200})S_{\rm NFW}}{\pi\ln(17)}\frac{1}
{r_{200}^2+16(d_A\theta)^2}\nonumber\\
&\times&\max\left[e^{-2.377(d_A\theta/r_{200}-1)},1\right],
\end{eqnarray}
where $S_{\rm NFW}$ is the integral flux of the smooth NFW halo of the
cluster, $d_A$ is the angular diameter distance.

To study the $\gamma$-ray line, we need to take into account the
energy resolution of the detector. We use an energy dependent
resolution from \cite{2012PhRvD..86b2002A}, where the energy resolution 
was derived through Monte-Carlo simulations with an integral over the
angular acceptance. Then we use the binned likelihood analysis
tool as in Sec. II to derive the upper limits of DM annihilation
final state $\gamma\gamma$. The point sources of the second
Fermi-LAT catalog \cite{2012ApJS..199...31N} located in the ROI are 
modeled simultaneously, with the normalizations free. The normalizations
of the diffuse backgrounds are also left free in the likelihood
fit. Besides the constraint from individual galaxy clusters, we
also do a combined likelihood analysis of all the seven galaxy
clusters simultaneously to combine the statistical power of these
different target regions \cite{2012JCAP...01..042H}. Constraints on the 
DM annihilation cross section are shown in Figure \ref{fig:sv_cl}, for
the cases without (left) and with (right) substructures. It can be
seen that the constraints from galaxy clusters are quite weak and
are still consistent with the DM interpretation of the inner
Galaxy line emission, even for the case with significant
substructure boost.

\FIGURE{
\includegraphics[width=0.45\columnwidth]{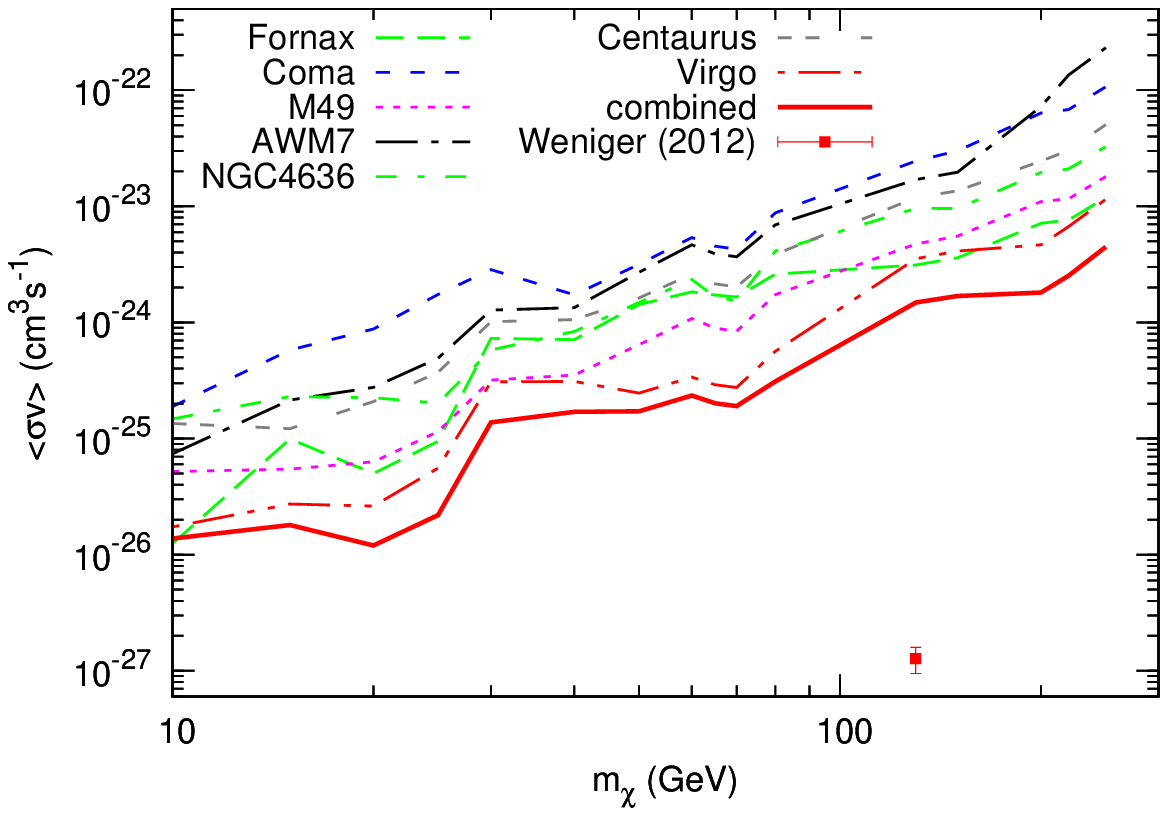}
\includegraphics[width=0.45\columnwidth]{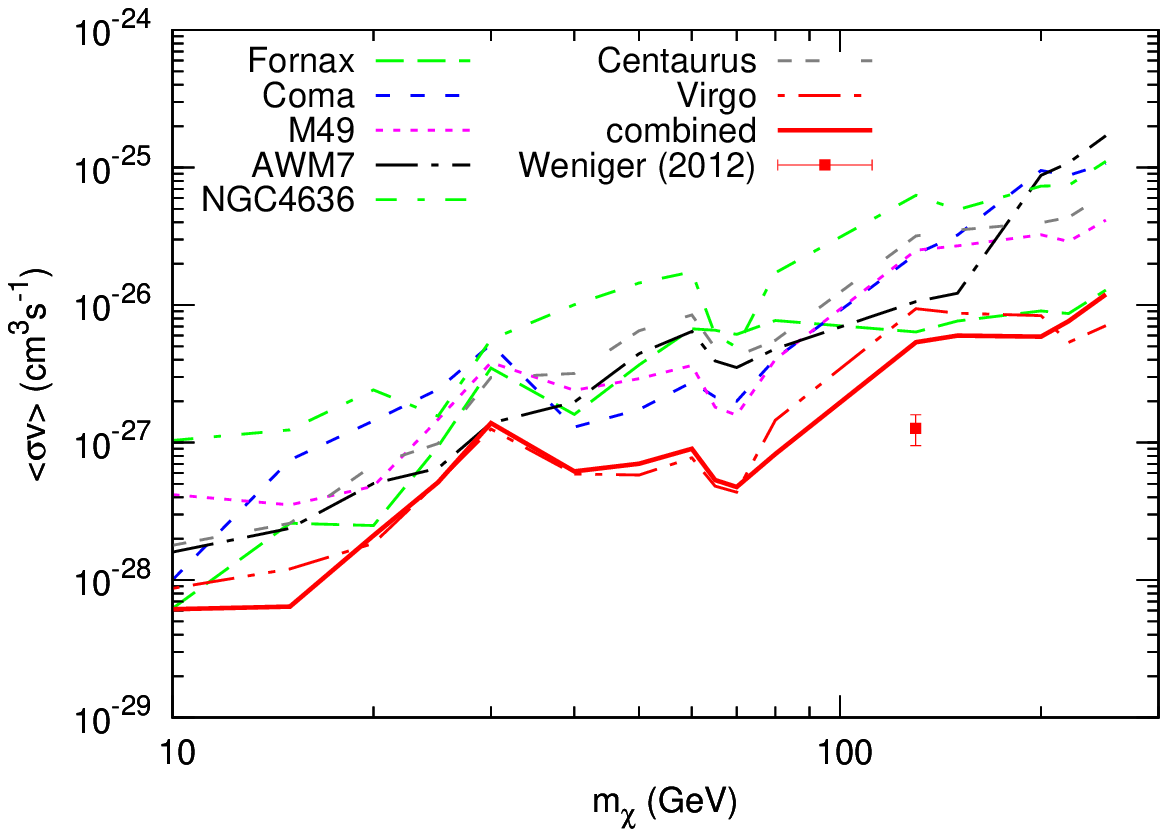}
\caption{Constraints on DM annihilation cross section to $\gamma\gamma$
from clusters without (left) and with (right) substructure enhancement.
\label{fig:sv_cl}}
}

\section{Line emission from dwarf galaxies}

Same as the analysis of galaxy clusters, we perform the search for
line emission with Fermi-LAT data of dwarf galaxies in this section.
In Ref. \cite{2012PhRvD..86b1302G} the authors have done a similar 
analysis and found no signal of line emission from the dwarf galaxies. 
Here we redo the analysis with a slightly different 
way\footnote{Specifically, our method is the same as the analysis of 
galaxy clusters, with point sources in the ROI and diffuse backgrounds 
involved.} from that of Ref. \cite{2012PhRvD..86b1302G}, for the seek 
of completeness and independent check. We use the same data selection 
conditions as above in the galaxy clusters analysis. The targets
chosen are Bootes I, Draco, Fornax, Sculptor, Segue 1, Sextans and
Ursa Minor, which are the same as in Ref. \cite{2012PhRvD..86b1302G}. 
For each dwarf galaxy, we model it as point source and the annihilation 
$J$-factor is adopted to be the best fit value given in 
\cite{2011PhRvL.107x1302A}. Then we perform the binned likelihood 
analysis to search for the line emission and derive the constraints 
on monochromatic $\gamma$-ray line flux.

\FIGURE{
\includegraphics[width=0.6\columnwidth]{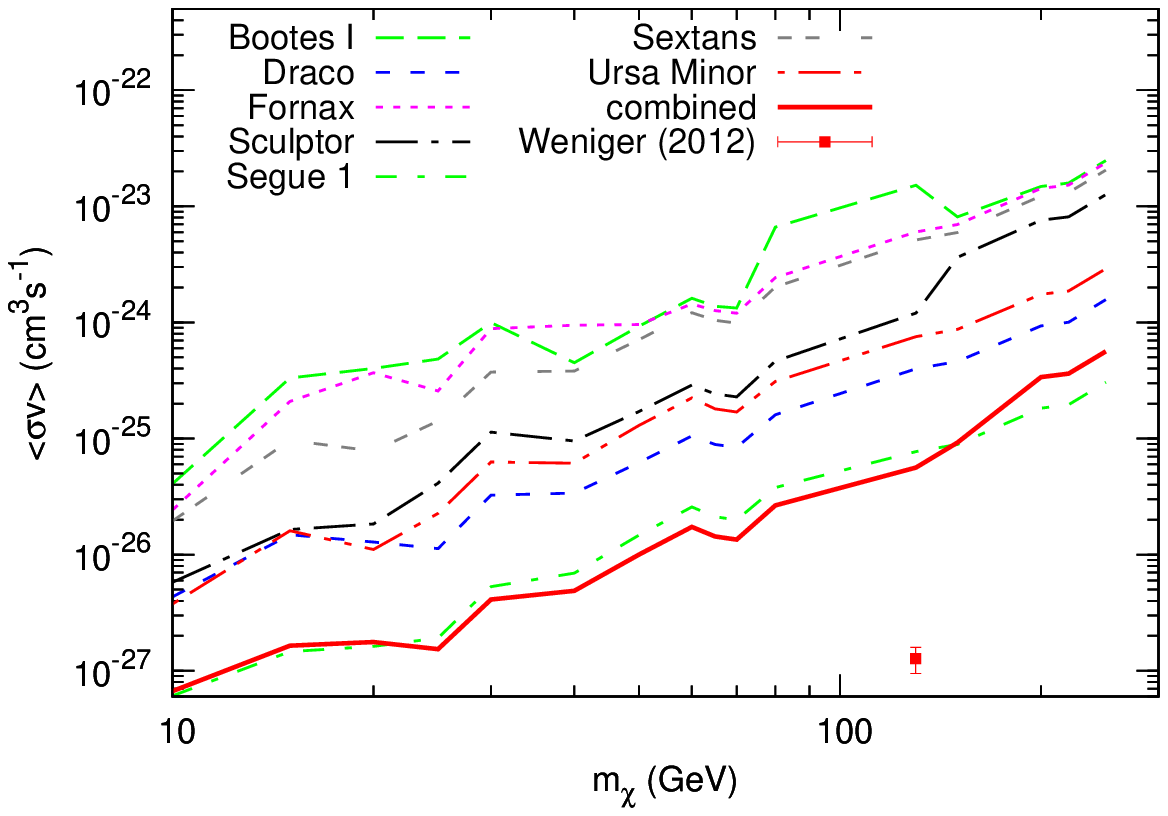}
\caption{Constraints on DM annihilation cross section to $\gamma\gamma$
from dwarf galaxies.
\label{fig:sv_dw}}
}

We have found no significant $\gamma$-ray line emission in the target
regions. The constraints on $\langle\sigma v\rangle_{\chi\chi\to\gamma
\gamma}$ are shown in Figure \ref{fig:sv_dw}. We can see that the upper
limits are still far away from the best fit point from Ref. 
\cite{2012JCAP...08..007W}. This result is consistent with the upper 
limit derived in \cite{2012PhRvD..86b1302G}, though we use a different 
method to deal with background photons. At the present time the 
constraint from dwarf galaxies can neither confirm nor exclude the 
DM-induced line interpretation of the Galactic center data.

\section{Conclusion and discussion}

The recently reported tentative $\gamma$-ray line emission from the
Fermi-LAT observations in the inner Galaxy \cite{2012JCAP...07..054B,
2012JCAP...08..007W,2012JCAP...09..032T,2012arXiv1205.4700B,
2012arXiv1206.1616S} has invoked many discussions of the DM
annihilation signals. In this paper we do a comprehensive analysis
using the Fermi-LAT data in the Galactic center region, Galactic halo,
galaxy clusters and dwarf galaxies, to test the DM annihilation
interpretation of this line emission.

Using the Fermi-LAT data in the Galactic center region, we get
strong constraints on the continuous $\gamma$-ray emission of DM
annihilation to $W^+W^-$, $b\bar{b}$, $\mu^+\mu^-$ and
$\tau^+\tau^-$ final states, which set useful constraints on DM
models to explain the $130$ GeV line emission.

We further perform the search for line emission with Fermi-LAT
data in the Milky Way halo. The constraints on the line emission 
are generally consistent with the tentative ``signal'' found in the 
inner Galaxy \cite{2012JCAP...08..007W}. Only when the enhancement effect 
due to DM substructures is taken into account, and the ``CLEAN'' events 
are studied, the constraints seem to have a weak tension with the DM 
annihilation interpretation of the $130$ GeV line emission if the DM 
density profile follows Einasto or NFW profiles. However, considering 
the uncertainties of both the constraints and the fitting ``signal'',
we can not draw a firm conclusion on it at present. We just mention 
that if it is finally proven to be true, one may need to assume cuspier 
density profile of DM in the Galactic center (e.g., accreted by the 
central supermassive black hole \cite{1999PhRvL..83.1719G}) to consistently 
understand the results in the inner Galaxy and the Galactic halo within 
DM annihilation scenario.

Galaxy clusters and dwarf galaxies are also studied to further
test the DM interpretation of the $130$ GeV line emission. It is
interesting to find that there is a concentration of photons in
$120-140$ GeV from the six nearby clusters (Figure 4) for the
``CLEAN'' and ``ULTRACLEAN'' classes of events. However, the
result seems hard to indicate an excess of line emission.
Therefore we set upper limits on the DM annihilation cross section
to monochromatic $\gamma$-ray line, which are consistent with the
claimed results in the inner Galaxy \cite{2012JCAP...08..007W}, even 
when the large boost factor of substructures are taken into account. 
No signal of line emission from dwarf galaxies is found, and the
constraints on DM annihilation cross section to monochromatic
$\gamma$-ray line are derived. Compared with the galaxy clusters,
the constraints are weaker for dwarf galaxies, and are consistent
with the claimed results in the inner Galaxy \cite{2012JCAP...08..007W}.

\acknowledgments

We thank Yi-Zhong Fan, Meng Su and Christoph Weniger for helpful 
discussion and useful comments on the manuscript.
This work is supported by National Natural Science Foundation of China
under grant Nos. 11075169, 11073024, 11105155, 11105157, the 973 project 
under grant No. 2010CB833000, and Chinese Academy of Sciences under grant 
No. KJCX2-EW-W01. QY acknowledges the support from the Key Laboratory 
of Dark Matter and Space Astronomy of Chinese Academy of Sciences. 
XLC acknowledges the support from John Templeton Foundation \& NAOC 
Beyond the Horizon program.

\bibliographystyle{JHEP}
\bibliography{/home/yuanq/work/cygnus/tex/refs}

\providecommand{\href}[2]{#2}\begingroup\raggedright\begin{thebibliography}{10}

\bibitem{2012JCAP...07..054B}
T.~{Bringmann}, X.~{Huang}, A.~{Ibarra}, S.~{Vogl}, and C.~{Weniger}, {\it
  {Fermi LAT search for internal bremsstrahlung signatures from dark matter
  annihilation}},  {\em \jcap} {\bf 7} (2012) 54
  [\href{http://arXiv.org/abs/arXiv/1203.1312}{{\tt arXiv:1203.1312}}].

\bibitem{2012JCAP...08..007W}
C.~{Weniger}, {\it {A tentative gamma-ray line from Dark Matter annihilation at
  the Fermi Large Area Telescope}},  {\em \jcap} {\bf 8} (2012) 7
  [\href{http://arXiv.org/abs/arXiv/1204.2797}{{\tt arXiv:1204.2797}}].

\bibitem{2012JCAP...09..032T}
E.~{Tempel}, A.~{Hektor}, and M.~{Raidal}, {\it {Fermi 130 GeV gamma-ray excess
  and dark matter annihilation in sub-haloes and in the Galactic centre}},
  {\em \jcap} {\bf 9} (2012) 32
  [\href{http://arXiv.org/abs/arXiv/1205.1045}{{\tt arXiv:1205.1045}}].

\bibitem{2012arXiv1205.4700B}
A.~{Boyarsky}, D.~{Malyshev}, and O.~{Ruchayskiy}, {\it {Spectral and spatial
  variations of the diffuse gamma-ray background in the vicinity of the
  Galactic plane and possible nature of the feature at 130 GeV}},  {\em ArXiv
  e-prints} (2012) [\href{http://arXiv.org/abs/arXiv/1205.4700}{{\tt
  arXiv:1205.4700}}].

\bibitem{2012arXiv1206.1616S}
M.~{Su} and D.~P. {Finkbeiner}, {\it {Strong Evidence for Gamma-ray Lines from
  the Inner Galaxy}},  {\em ArXiv e-prints} (2012)
  [\href{http://arXiv.org/abs/arXiv/1206.1616}{{\tt arXiv:1206.1616}}].

\bibitem{2012arXiv1205.1520D}
E.~{Dudas}, Y.~{Mambrini}, S.~{Pokorski}, and A.~{Romagnoni}, {\it {Extra U(1)
  as natural source of a monochromatic gamma ray line}},  {\em ArXiv e-prints}
  (2012) [\href{http://arXiv.org/abs/arXiv/1205.1520}{{\tt arXiv:1205.1520}}].

\bibitem{2012PhRvD..86a5016C}
J.~M. {Cline}, {\it {130 GeV dark matter and the Fermi gamma-ray line}},  {\em
  \prd} {\bf 86} (2012) 015016
  [\href{http://arXiv.org/abs/arXiv/1205.2688}{{\tt arXiv:1205.2688}}].

\bibitem{2012PhRvD..86d3515C}
K.-Y. {Choi} and O.~{Seto}, {\it {Dirac right-handed sneutrino dark matter and
  its signature in the gamma-ray lines}},  {\em \prd} {\bf 86} (2012) 043515
  [\href{http://arXiv.org/abs/arXiv/1205.3276}{{\tt arXiv:1205.3276}}].

\bibitem{2012arXiv1205.4151K}
B.~{Kyae} and J.-C. {Park}, {\it {130 GeV Gamma-Ray Line from Dark Matter
  Decay}},  {\em ArXiv e-prints} (2012)
  [\href{http://arXiv.org/abs/arXiv/1205.4151}{{\tt arXiv:1205.4151}}].

\bibitem{2012arXiv1205.4675L}
H.~M. {Lee}, M.~{Park}, and W.-I. {Park}, {\it {Fermi Gamma Ray Line at 130 GeV
  from Axion-Mediated Dark Matter}},  {\em ArXiv e-prints} (2012)
  [\href{http://arXiv.org/abs/arXiv/1205.4675}{{\tt arXiv:1205.4675}}].

\bibitem{2012JCAP...09..003R}
A.~{Rajaraman}, T.~M.~P. {Tait}, and D.~{Whiteson}, {\it {Two lines or not two
  lines? That is the question of gamma ray spectra}},  {\em \jcap} {\bf 9}
  (2012) 3 [\href{http://arXiv.org/abs/arXiv/1205.4723}{{\tt
  arXiv:1205.4723}}].

\bibitem{2012arXiv1205.5789S}
B.~{Samir Acharya}, G.~{Kane}, P.~{Kumar}, R.~{Lu}, and B.~{Zheng}, {\it {Mixed
  Wino-Axion Dark Matter in String/M Theory and the 130 GeV Gamma-line
  ''Signal''}},  {\em ArXiv e-prints} (2012)
  [\href{http://arXiv.org/abs/arXiv/1205.5789}{{\tt arXiv:1205.5789}}].

\bibitem{2012PhRvD..86h3521C}
X.~{Chu}, T.~{Hambye}, T.~{Scarna}, and M.~H.~G. {Tytgat}, {\it {What if dark
  matter gamma-ray lines come with gluon lines?}},  {\em \prd} {\bf 86} (2012)
  083521 [\href{http://arXiv.org/abs/arXiv/1206.2279}{{\tt arXiv:1206.2279}}].

\bibitem{2012arXiv1206.2910W}
N.~{Weiner} and I.~{Yavin}, {\it {How Dark Are Majorana WIMPs? Signals from
  MiDM and Rayleigh Dark Matter}},  {\em ArXiv e-prints} (2012)
  [\href{http://arXiv.org/abs/arXiv/1206.2910}{{\tt arXiv:1206.2910}}].

\bibitem{2012arXiv1206.4758F}
L.~{Feng}, Q.~{Yuan}, and Y.-Z. {Fan}, {\it {Tentative wiggle in the cosmic ray
  electron/positron spectrum at \$$\backslash$sim\$100 GeV: a dark matter
  annihilation signal in accordance with the 130 GeV \$$\backslash$gamma-\$ray
  line?}},  {\em ArXiv e-prints} (2012)
  [\href{http://arXiv.org/abs/arXiv/1206.4758}{{\tt arXiv:1206.4758}}].

\bibitem{2012arXiv1207.1341H}
J.~H. {Heo} and C.~S. {Kim}, {\it {Cosmic ray signatures of Dipole-Interacting
  Fermionic Dark Matter}},  {\em ArXiv e-prints} (2012)
  [\href{http://arXiv.org/abs/arXiv/1207.1341}{{\tt arXiv:1207.1341}}].

\bibitem{2012arXiv1207.1537O}
I.~{Oda}, {\it {Remarks on Two Gamma Ray Lines from the Inner Galaxy}},  {\em
  ArXiv e-prints} (2012) [\href{http://arXiv.org/abs/arXiv/1207.1537}{{\tt
  arXiv:1207.1537}}].

\bibitem{2012PhLB..715..285Y}
R.-Z. {Yang}, Q.~{Yuan}, L.~{Feng}, Y.-Z. {Fan}, and J.~{Chang}, {\it
  {Statistical interpretation of the spatial distribution of current 130 GeV
  {$\gamma$}-ray line signal within the dark matter annihilation scenario}},
  {\em Physics Letters B} {\bf 715} (2012) 285--288
  [\href{http://arXiv.org/abs/arXiv/1207.1621}{{\tt arXiv:1207.1621}}].

\bibitem{2012JCAP...10..033F}
M.~T. {Frandsen}, U.~{Haisch}, F.~{Kahlhoefer}, P.~{Mertsch}, and
  K.~{Schmidt-Hoberg}, {\it {Loop-induced dark matter direct detection signals
  from {$\gamma$}-ray lines}},  {\em \jcap} {\bf 10} (2012) 33
  [\href{http://arXiv.org/abs/arXiv/1207.3971}{{\tt arXiv:1207.3971}}].

\bibitem{2012arXiv1207.4981P}
J.-C. {Park} and S.~C. {Park}, {\it {Radiatively decaying scalar dark matter
  through U(1) mixings and the Fermi 130 GeV gamma-ray line}},  {\em ArXiv
  e-prints} (2012) [\href{http://arXiv.org/abs/arXiv/1207.4981}{{\tt
  arXiv:1207.4981}}].

\bibitem{2012JCAP...08..003D}
D.~{Das}, U.~{Ellwanger}, and P.~{Mitropoulos}, {\it {A 130 GeV photon line
  from dark matter annihilation in the NMSSM}},  {\em \jcap} {\bf 8} (2012) 3
  [\href{http://arXiv.org/abs/arXiv/1206.2639}{{\tt arXiv:1206.2639}}].

\bibitem{2012arXiv1206.2863K}
Z.~{Kang}, T.~{Li}, J.~{Li}, and Y.~{Liu}, {\it {Brightening the (130 GeV)
  Gamma-Ray Line}},  {\em ArXiv e-prints} (2012)
  [\href{http://arXiv.org/abs/arXiv/1206.2863}{{\tt arXiv:1206.2863}}].

\bibitem{2012JCAP...07..011P}
S.~{Profumo} and T.~{Linden}, {\it {Gamma-ray lines in the Fermi data: is it a
  bubble?}},  {\em \jcap} {\bf 7} (2012) 11
  [\href{http://arXiv.org/abs/arXiv/1204.6047}{{\tt arXiv:1204.6047}}].

\bibitem{2012arXiv1207.0458A}
F.~{Aharonian}, D.~{Khangulyan}, and D.~{Malyshev}, {\it {Cold
  ultrarelativistic pulsar winds as potential sources of galactic gamma-ray
  lines above 100 GeV}},  {\em ArXiv e-prints} (2012)
  [\href{http://arXiv.org/abs/arXiv/1207.0458}{{\tt arXiv:1207.0458}}].

\bibitem{2012PhRvD..86d3524B}
M.~R. {Buckley} and D.~{Hooper}, {\it {Implications of a 130 GeV gamma-ray line
  for dark matter}},  {\em \prd} {\bf 86} (2012) 043524
  [\href{http://arXiv.org/abs/arXiv/1205.6811}{{\tt arXiv:1205.6811}}].

\bibitem{2012arXiv1206.7056B}
W.~{Buchmuller} and M.~{Garny}, {\it {Decaying vs Annihilating Dark Matter in
  Light of a Tentative Gamma-Ray Line}},  {\em ArXiv e-prints} (2012)
  [\href{http://arXiv.org/abs/arXiv/1206.7056}{{\tt arXiv:1206.7056}}].

\bibitem{2012arXiv1207.0800C}
T.~{Cohen}, M.~{Lisanti}, T.~R. {Slatyer}, and J.~G. {Wacker}, {\it
  {Illuminating the 130 GeV Gamma Line with Continuum Photons}},  {\em ArXiv
  e-prints} (2012) [\href{http://arXiv.org/abs/arXiv/1207.0800}{{\tt
  arXiv:1207.0800}}].

\bibitem{2012PhRvD..86h3525C}
I.~{Cholis}, M.~{Tavakoli}, and P.~{Ullio}, {\it {Searching for the continuum
  spectrum photons correlated to the 130 GeV gamma-ray line}},  {\em \prd} {\bf
  86} (2012) 083525 [\href{http://arXiv.org/abs/arXiv/1207.1468}{{\tt
  arXiv:1207.1468}}].

\bibitem{2012PhRvD..86b1302G}
A.~{Geringer-Sameth} and S.~M. {Koushiappas}, {\it {Dark matter line search
  using a joint analysis of dwarf galaxies with the Fermi Gamma-ray Space
  Telescope}},  {\em \prd} {\bf 86} (2012) 021302
  [\href{http://arXiv.org/abs/arXiv/1206.0796}{{\tt arXiv:1206.0796}}].

\bibitem{2012arXiv1207.4466H}
A.~{Hektor}, M.~{Raidal}, and E.~{Tempel}, {\it {An evidence for indirect
  detection of dark matter from galaxy clusters in Fermi-LAT data}},  {\em
  ArXiv e-prints} (2012) [\href{http://arXiv.org/abs/arXiv/1207.4466}{{\tt
  arXiv:1207.4466}}].

\bibitem{2012arXiv1207.7060S}
M.~{Su} and D.~P. {Finkbeiner}, {\it {Double Gamma-ray Lines from Unassociated
  Fermi-LAT Sources}},  {\em ArXiv e-prints} (2012)
  [\href{http://arXiv.org/abs/arXiv/1207.7060}{{\tt arXiv:1207.7060}}].

\bibitem{2012PhLB..715...35L}
Y.~{Li} and Q.~{Yuan}, {\it {Testing the 130 GeV gamma-ray line with high
  energy resolution detectors}},  {\em Physics Letters B} {\bf 715} (2012)
  35--37 [\href{http://arXiv.org/abs/arXiv/1206.2241}{{\tt arXiv:1206.2241}}].

\bibitem{2012arXiv1207.6773B}
L.~{Bergstr{\"o}m}, G.~{Bertone}, J.~{Conrad}, C.~{Farnier}, and C.~{Weniger},
  {\it {Investigating Gamma-Ray Lines from Dark Matter with Future
  Observatories}},  {\em ArXiv e-prints} (2012)
  [\href{http://arXiv.org/abs/arXiv/1207.6773}{{\tt arXiv:1207.6773}}].

\bibitem{2012PhRvD..86b2002A}
M.~{Ackermann}, M.~{Ajello}, A.~{Albert}, {et~al.}, {\it {Fermi LAT search for
  dark matter in gamma-ray lines and the inclusive photon spectrum}},  {\em
  \prd} {\bf 86} (2012) 022002
  [\href{http://arXiv.org/abs/arXiv/1205.2739}{{\tt arXiv:1205.2739}}].

\bibitem{2012MNRAS.419.1721G}
L.~{Gao}, C.~S. {Frenk}, A.~{Jenkins}, V.~{Springel}, and S.~D.~M. {White},
  {\it {Where will supersymmetric dark matter first be seen?}},  {\em \mnras}
  {\bf 419} (2012) 1721--1726 [\href{http://arXiv.org/abs/arXiv/1107.1916}{{\tt
  arXiv:1107.1916}}].

\bibitem{2012MNRAS.425.2169G}
L.~{Gao}, J.~F. {Navarro}, C.~S. {Frenk}, {et~al.}, {\it {The Phoenix Project:
  the dark side of rich Galaxy clusters}},  {\em \mnras} {\bf 425} (2012)
  2169--2186 [\href{http://arXiv.org/abs/arXiv/1201.1940}{{\tt
  arXiv:1201.1940}}].

\bibitem{2011PhRvL.107x1302A}
M.~{Ackermann}, M.~{Ajello}, A.~{Albert}, {et~al.}, {\it {Constraining Dark
  Matter Models from a Combined Analysis of Milky Way Satellites with the Fermi
  Large Area Telescope}},  {\em Physical Review Letters} {\bf 107} (2011)
  241302 [\href{http://arXiv.org/abs/arXiv/1108.3546}{{\tt arXiv:1108.3546}}].

\bibitem{1998ApJ...509..212S}
A.~W. {Strong} and I.~V. {Moskalenko}, {\it {Propagation of Cosmic-Ray Nucleons
  in the Galaxy}},  {\em \apj} {\bf 509} (1998) 212--228
  [\href{http://arXiv.org/abs/arXiv:astro-ph/9807150}{{\tt
  arXiv:astro-ph/9807150}}].

\bibitem{2012ApJS..199...31N}
P.~L. {Nolan}, A.~A. {Abdo}, M.~{Ackermann}, {et~al.}, {\it {Fermi Large Area
  Telescope Second Source Catalog}},  {\em \apjs} {\bf 199} (2012) 31
  [\href{http://arXiv.org/abs/arXiv/1108.1435}{{\tt arXiv:1108.1435}}].

\bibitem{2006JHEP...05..026S}
T.~{Sj{\"o}strand}, S.~{Mrenna}, and P.~{Skands}, {\it {PYTHIA 6.4 physics and
  manual}},  {\em Journal of High Energy Physics} {\bf 5} (2006) 26
  [\href{http://arXiv.org/abs/arXiv:hep-ph/0603175}{{\tt
  arXiv:hep-ph/0603175}}].

\bibitem{1997ApJ...490..493N}
J.~F. {Navarro}, C.~S. {Frenk}, and S.~D.~M. {White}, {\it {A Universal Density
  Profile from Hierarchical Clustering}},  {\em \apj} {\bf 490} (1997) 493
  [\href{http://arXiv.org/abs/arXiv:astro-ph/9611107}{{\tt
  arXiv:astro-ph/9611107}}].

\bibitem{1965TrAlm...5...87E}
J.~{Einasto}, {\it {On the Construction of a Composite Model for the Galaxy and
  on the Determination of the System of Galactic Parameters}},  {\em Trudy
  Astrofizicheskogo Instituta Alma-Ata} {\bf 5} (1965) 87--100.

\bibitem{2010JCAP...08..004C}
R.~{Catena} and P.~{Ullio}, {\it {A novel determination of the local dark
  matter density}},  {\em \jcap} {\bf 8} (2010) 4
  [\href{http://arXiv.org/abs/arXiv/0907.0018}{{\tt arXiv:0907.0018}}].

\bibitem{2008MNRAS.391.1685S}
V.~{Springel}, J.~{Wang}, M.~{Vogelsberger}, {et~al.}, {\it {The Aquarius
  Project: the subhaloes of galactic haloes}},  {\em \mnras} {\bf 391} (2008)
  1685--1711 [\href{http://arXiv.org/abs/arXiv/0809.0898}{{\tt
  arXiv:0809.0898}}].

\bibitem{2012arXiv1203.5636Y}
Q.~{Yuan}, Y.~{Cao}, J.~{Liu}, {et~al.}, {\it {Gamma-rays From Warm WIMP Dark
  Matter Annihilation}},  {\em ArXiv e-prints} (2012)
  [\href{http://arXiv.org/abs/arXiv/1203.5636}{{\tt arXiv:1203.5636}}].

\bibitem{2011PhRvD..84l3005H}
D.~{Hooper} and T.~{Linden}, {\it {Origin of the gamma rays from the Galactic
  Center}},  {\em \prd} {\bf 84} (2011) 123005
  [\href{http://arXiv.org/abs/arXiv/1110.0006}{{\tt arXiv:1110.0006}}].

\bibitem{2012ApJS..203....4A}
M.~{Ackermann}, M.~{Ajello}, A.~{Albert}, {et~al.}, {\it {The Fermi Large Area
  Telescope on Orbit: Event Classification, Instrument Response Functions, and
  Calibration}},  {\em \apjs} {\bf 203} (2012) 4
  [\href{http://arXiv.org/abs/arXiv/1206.1896}{{\tt arXiv:1206.1896}}].

\bibitem{1994APh.....2..261B}
L.~{Bergstr{\"o}m} and J.~{Kaplan}, {\it {Gamma ray lines from TeV dark
  matter}},  {\em Astroparticle Physics} {\bf 2} (1994) 261--268
  [\href{http://arXiv.org/abs/arXiv:hep-ph/9403239}{{\tt
  arXiv:hep-ph/9403239}}].

\bibitem{1995PhRvD..51.3121J}
G.~{Jungman} and M.~{Kamionkowski}, {\it {{$\gamma$} rays from neutralino
  annihilation}},  {\em \prd} {\bf 51} (1995) 3121--3124
  [\href{http://arXiv.org/abs/arXiv:hep-ph/9501365}{{\tt
  arXiv:hep-ph/9501365}}].

\bibitem{1997PhLB..411...86B}
Z.~{Bern}, P.~{Gondolo}, and M.~{Perelstein}, {\it {Neutralino annihilation
  into two photons}},  {\em Physics Letters B} {\bf 411} (1997) 86--96
  [\href{http://arXiv.org/abs/arXiv:hep-ph/9706538}{{\tt
  arXiv:hep-ph/9706538}}].

\bibitem{1997NuPhB.504...27B}
L.~{Bergstr{\"o}m} and P.~{Ullio}, {\it {Full one-loop calculation of
  neutralino annihilation into two photons}},  {\em Nuclear Physics B} {\bf
  504} (1997) 27--44 [\href{http://arXiv.org/abs/arXiv:hep-ph/9706232}{{\tt
  arXiv:hep-ph/9706232}}].

\bibitem{2005NIMPA.551..493R}
W.~A. {Rolke}, A.~M. {L{\'o}pez}, and J.~{Conrad}, {\it {Limits and confidence
  intervals in the presence of nuisance parameters}},  {\em Nuclear Instruments
  and Methods in Physics Research A} {\bf 551} (2005) 493--503
  [\href{http://arXiv.org/abs/arXiv:physics/0403059}{{\tt
  arXiv:physics/0403059}}].

\bibitem{2011PhRvD..84l3509P}
A.~{Pinzke}, C.~{Pfrommer}, and L.~{Bergstr{\"o}m}, {\it {Prospects of
  detecting gamma-ray emission from galaxy clusters: Cosmic rays and dark
  matter annihilations}},  {\em \prd} {\bf 84} (2011) 123509
  [\href{http://arXiv.org/abs/arXiv/1105.3240}{{\tt arXiv:1105.3240}}].

\bibitem{2002ApJ...567..716R}
T.~H. {Reiprich} and H.~{B{\"o}hringer}, {\it {The Mass Function of an X-Ray
  Flux-limited Sample of Galaxy Clusters}},  {\em \apj} {\bf 567} (2002)
  716--740 [\href{http://arXiv.org/abs/arXiv:astro-ph/0111285}{{\tt
  arXiv:astro-ph/0111285}}].

\bibitem{2007A&A...466..805C}
Y.~{Chen}, T.~H. {Reiprich}, H.~{B{\"o}hringer}, Y.~{Ikebe}, and Y.-Y. {Zhang},
  {\it {Statistics of X-ray observables for the cooling-core and non-cooling
  core galaxy clusters}},  {\em \aap} {\bf 466} (2007) 805--812
  [\href{http://arXiv.org/abs/arXiv:astro-ph/0702482}{{\tt
  arXiv:astro-ph/0702482}}].

\bibitem{2012JCAP...01..042H}
X.~{Huang}, G.~{Vertongen}, and C.~{Weniger}, {\it {Probing dark matter decay
  and annihilation with Fermi LAT observations of nearby galaxy clusters}},
  {\em \jcap} {\bf 1} (2012) 42
  [\href{http://arXiv.org/abs/arXiv/1110.1529}{{\tt arXiv:1110.1529}}].

\bibitem{2008MNRAS.390L..64D}
A.~R. {Duffy}, J.~{Schaye}, S.~T. {Kay}, and C.~{Dalla Vecchia}, {\it {Dark
  matter halo concentrations in the Wilkinson Microwave Anisotropy Probe year 5
  cosmology}},  {\em \mnras} {\bf 390} (2008) L64--L68
  [\href{http://arXiv.org/abs/arXiv/0804.2486}{{\tt arXiv:0804.2486}}].

\bibitem{2012arXiv1201.1003H}
J.~{Han}, C.~S. {Frenk}, V.~R. {Eke}, L.~{Gao}, and S.~D.~M. {White}, {\it
  {Evidence for extended gamma-ray emission from galaxy clusters}},  {\em ArXiv
  e-prints} (2012) [\href{http://arXiv.org/abs/arXiv/1201.1003}{{\tt
  arXiv:1201.1003}}].

\bibitem{1999PhRvL..83.1719G}
P.~{Gondolo} and J.~{Silk}, {\it {Dark Matter Annihilation at the Galactic
  Center}},  {\em Physical Review Letters} {\bf 83} (1999) 1719--1722
  [\href{http://arXiv.org/abs/arXiv:astro-ph/9906391}{{\tt
  arXiv:astro-ph/9906391}}].

\end{thebibliography}\endgroup

\end{document}